\documentclass{article}

     \usepackage[preprint,nonatbib]{neurips_2023}

\usepackage[utf8]{inputenc} % allow utf-8 input
\usepackage[T1]{fontenc}    % use 8-bit T1 fonts
\usepackage{hyperref}       % hyperlinks
\usepackage{url}            % simple URL typesetting
\usepackage{booktabs}       % professional-quality tables
\usepackage{amsfonts}       % blackboard math symbols
\usepackage{nicefrac}       % compact symbols for 1/2, etc.
\usepackage{microtype}      % microtypography
\usepackage{xcolor}         % colors
\usepackage{graphicx}  

\title{On the Emergence of Symmetrical Reality}

\author{%
 \hspace{-0.9cm} Zhenliang Zhang$^{1}$\thanks{Corresponding authors.}\ , Zeyu Zhang$^{1}$,  Ziyuan Jiao$^{1}$,  Yao Su$^{1}$,  Hangxin Liu$^{1}$,  Wei Wang$^{1}$,  Song-Chun Zhu$^{1,2*}$ \\
\hspace{-0.9cm}  $^{1}$ National Key Laboratory of General Artificial Intelligence, BIGAI\\%
\hspace{-0.9cm}  $^{2}$ School of Intelligence Science and Technology, Peking University \\%
\hspace{-0.9cm}  \texttt{\{zlzhang, zhangzeyu, jiaoziyuan, suyao, liuhx, wangwei, sczhu\}@bigai.ai} \\
}

\newcommand{\ie}{\textit{i.e.}}
\newcommand{\eg}{\textit{e.g.}}
\newcommand{\etal}{\textit{et~al.}}
\newcommand{\etc}{\textit{etc}}

\begin{document}

\maketitle

\begin{abstract}
Artificial intelligence (AI) has revolutionized human cognitive abilities and facilitated the development of new AI entities capable of interacting with humans in both physical and virtual environments. Despite the existence of virtual reality, mixed reality, and augmented reality for several years, integrating these technical fields remains a formidable challenge due to their disparate application directions. The advent of AI agents, capable of autonomous perception and action, further compounds this issue by exposing the limitations of traditional human-centered research approaches. It is imperative to establish a comprehensive framework that accommodates the dual perceptual centers of humans and AI agents in both physical and virtual worlds. In this paper, we introduce the symmetrical reality framework, which offers a unified representation encompassing various forms of physical-virtual amalgamations. This framework enables researchers to better comprehend how AI agents can collaborate with humans and how distinct technical pathways of physical-virtual integration can be consolidated from a broader perspective. We then delve into the coexistence of humans and AI, demonstrating a prototype system that exemplifies the operation of symmetrical reality systems for specific tasks, such as pouring water. Subsequently, we propose an instance of an AI-driven active assistance service that illustrates the potential applications of symmetrical reality. This paper aims to offer beneficial perspectives and guidance for researchers and practitioners in different fields, thus contributing to the ongoing research about human-AI coexistence in both physical and virtual environments.
\end{abstract}

\section{Introduction}

\begin{figure}
  \centering
  \includegraphics[width=\linewidth]{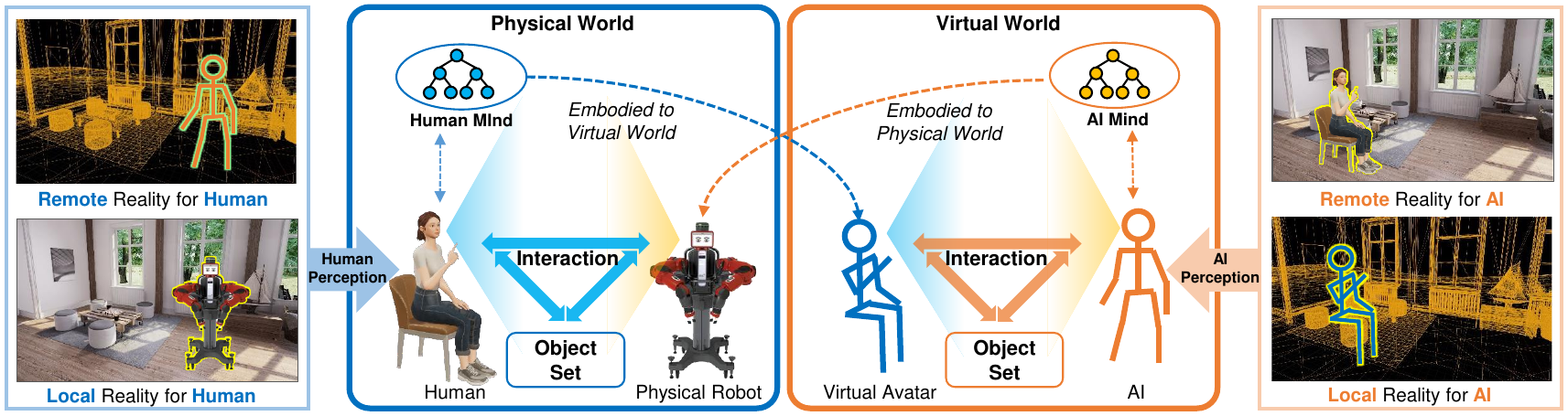}
\caption{
The topological structure of symmetrical reality. Symmetrical reality integrates the physical and virtual worlds into a unified system. Humans inhabit the physical world and possess virtual embodiments within the virtual domain, while AI agents originate in the virtual world and maintain physical embodiments in the physical world. The symmetrical attributes of this system are manifested through symmetrical perception and bidirectional interaction between humans and AI agents.
 }
  \label{fig:teaser}
\end{figure}

Artificial intelligence (AI) has experienced exponential growth in numerous sectors of contemporary society~\cite{nadikattu2016emerging,edmonds2019tale}, with groundbreaking research~\cite{lu2018brain,li2020artificial} catalyzing significant transformations in daily life. The intelligence of AI agents~\cite{park2023generative,wang2023voyager} is rapidly advancing with the emergence of new technologies. For example, large language models such as ChatGPT~\cite{zhuo2023exploring} have completed tasks designed to assess the ``Theory of Mind'' (ToM)~\cite{kosinski2023theory} to a large extent. Notably, the recently developed GPT-4~\cite{openai2023gpt4} has even passed bar exams intended for humans~\cite{katz2023gpt}, regardless of whether it possesses the necessary knowledge and reasoning ability. These sophisticated AI models hold the potential to revolutionize human-AI interactions in both physical and virtual environments.

The ability of artificial intelligence to generate intelligent agents, which makes the agent exhibit perceptual capabilities comparable to humans~\cite{zhang2022towards} or smart behaviors based on perceptual computing~\cite{sheth2016internet}, draws people's attention to the perceptual processes of these AI agents. This form of perception, which operates in a direction inverse to human perception, can give rise to symmetrical reality (SR)~\cite{zhang2019symmetrical} within the context of physical-virtual world integration. From this perspective, virtual reality (VR)~\cite{zhao2009survey,zhou2009virtual}, mixed reality (MR)~\cite{costanza2009mixed}, and augmented reality (AR)~\cite{azuma1997survey,van2010survey} represent specific amalgamations of physical and virtual worlds that consider only one perceptual center (\ie, physical humans) while disregarding the other perceptual center (\ie, AI agents) situated in the virtual world. In other words, VR, MR, and AR can be viewed as particular instances of symmetrical reality.

The theoretical framework of symmetrical reality holds significant value for three primary reasons. Firstly, there currently exists no comprehensive framework that fully encompasses the entire domain of physical-virtual world integration. For instance, existing forms such as VR and AR are typically presented as distinct application cases, despite the potential for their fusion~\cite{lifton2009dual}. Secondly, the incorporation of AI agents increases the complexity of system structures, which cannot be adequately described by current VR and AR frameworks. Symmetrical reality, with its dual-center perception feature, is well-equipped to address these emerging scenarios. Thirdly, AI agents are fundamentally virtual entities (comprising algorithms or models) situated within an artificial virtual space, capable of concurrently observing and influencing both virtual and physical worlds. This complex feature can only be accurately represented within a symmetrical reality framework. To some extent, the advancement of AI has facilitated the emergence of symmetrical reality.

Symmetrical reality and AI are mutually reinforcing. On one hand, AI supports the symmetrical structure of symmetrical reality by providing the perceptual center of the virtual world (\ie, the AI agent). On the other hand, symmetrical reality fosters the advancement of AI by facilitating deep connections between humans and AI agents. Within a symmetrical reality system, the physical and virtual worlds are uniformly represented, with physical humans constituting components of the physical world and AI agents comprising components of the virtual world. Humans and AI agents coexist within this framework, enabling the exchange of information across physical and virtual worlds. This gives rise to the concept of human-compatible AI, as explored by Russell~\cite{russell2019human}, which addresses potential challenges in human-AI coexistence. According to his interpretation, human-compatible AI should remain under the control of human developers and serve as a beneficial asset to society, with human-AI cooperation and communication actively encouraged. The mentioned term ``under control'' should be understood as creating beneficial AI and instructing it to cooperate with humans, rather than directly controlling its thoughts and actions.

We build upon the aforementioned arguments and expand the scope of human-AI interaction to encompass scenarios that consider both physical and virtual spaces, with due regard for ethics and safety. Within the context of physical-virtual world integration, VR and AR frameworks designate humans as the perceptual center, while inverse virtual reality (IVR)~\cite{zhang2018ivr} and inverse augmented reality (IAR)~\cite{zhang2018iar} frameworks designate AI agents as the perceptual center. By facilitating information exchange across disparate worlds, symmetrical reality accelerates the development of human-compatible AI and significantly enhances human-AI interaction~\cite{jarrahi2018artificial}.

The forms of existence of AI and humans are fundamentally distinct. Humans are biological entities inhabiting the physical world, while AI is an artificial construct that does not naturally occur in nature (although physical embodiments of AI are permitted within the physical world). The divergent ontological positions~\cite{mantovani1999real} of AI and humans can give rise to distinct cognitive paradigms. Consequently, within the context of physical-virtual world integration, AI agents are virtual entities that serve as the perceptual center of the virtual world, while humans are physical entities that serve as the perceptual center of the physical world. Similar to human cognition, AI agents can also function as cognitive entities with specific cognitive approaches, but are not necessarily bound to adhere to human cognitive patterns. The significance of symmetrical reality lies in its capacity to comprehensively model human-AI systems and elucidate fundamental characteristics of human-AI interaction.

This paper presents a detailed exposition of the theoretical framework of symmetrical reality, as depicted in Fig.~\ref{fig:teaser}. Our contributions are threefold: 

(i) We introduce a comprehensive blueprint of symmetrical reality, encompassing its framework and technical challenges, and elucidate the emergence and significance of symmetrical reality in light of recent advancements in artificial general intelligence. 

(ii) We examine the long-term coexistence of humans and AI to reveal the implications of human-compatible AI, and employ the illustrative task of pouring water to demonstrate the complete architecture of symmetrical reality systems. 

(iii) We showcase two typical system designs along with an example of active helping service to demonstrate the applicability of symmetrical reality to daily life, and offer recommendations for its future development.

The remainder of this paper is structured as follows. Section~\ref{sec:relatedwork} reviews related work from various perspectives. Section~\ref{sec:framework} formally introduces the comprehensive framework of symmetrical reality. Section~\ref{sec:tcsr} outlines the technical challenges associated with symmetrical reality, while Section~\ref{sec:coexistence} further explores the mutual influence of humans and AI agents within this system during their coexistence, including an example of constructing symmetrical reality systems through agent simulation and bidirectional interaction. Section~\ref{sec:poc} presents several representative symmetrical reality system designs and application scenarios. The paper concludes with a discussion, followed by a summary and directions for future research.

\section{Related Work} \label{sec:relatedwork}

We have structured the related work according to the logical framework of symmetrical reality. Four key components (\ie, physical environments, physical humans, virtual environments, and AI agents) are essential, along with two interaction paradigms (\ie, physical-virtual interaction and human-AI interaction) and one theoretical framework. In this section, we provide a brief overview of five of these seven components, excluding the two components (\ie, physical environments and physical humans) that are already well-known.

\subsection{Virtual Environments}

Various devices facilitate the creation of distinct categories of virtual environments~\cite{benford1996shared, benford1998understanding}. Regardless of the specific device employed, virtual environments constitute a crucial component of the SR framework by providing task environments for AI agents. Physical and virtual elements can be integrated to construct mixed environments~\cite{gaschler2014intuitive} for general tasks, with such virtual environments often serving educational or training purposes~~\cite{johnson2017driving,das2018embodied,zhu2015understanding}. Additionally, these environments can supply knowledge and interactive scenarios for AI tasks, such as indoor navigation~\cite{zhu2017target} and teleoperation~~\cite{lipton2018baxter}.

In addition to the aforementioned customized environments, cooperative working environments~\cite{takemura1992cooperative, szalavari1998studierstube, greenhalgh1995massive, gaver1992realizing, fahlen1993space, bly1993media, benford1997crowded} have gained widespread popularity. When constructing virtual environments, certain typical devices are utilized. For instance, Ishii~\etal~introduced a shared drawing medium~\cite{ishii1993integration} and proposed ``Tangible Bits''~\cite{ishii1997tangible} to enhance interfaces. To display virtual information to users, video see-through head-mounted displays~\cite{edwards1993video,bajura1992merging} and optical see-through head-mounted displays~\cite{thomas1992augmented} are commonly employed due to their functional properties. While working in virtual environments~\cite{fitzpatrick1996physical} offers several advantages, balancing privacy and awareness remains a challenge~\cite{hudson1996techniques}.

\subsection{AI Agents}

AI agents are integral components of the SR framework, with their rapidly evolving perceptual and cognitive abilities exerting a significant influence on the trajectory of AI research. To comprehend current trends, we will briefly introduce two important concepts, \ie, cognitive AI and explainable AI.

Cognitive AI~\cite{salomon1988ai,forbus2006companion} supports humans in numerous capacities. To develop an AI agent into a cognitive AI, extensive research has been conducted in the field of embodied AI~\cite{duan2022survey}.
These tasks are typically accomplished through the construction of various virtual simulators~\cite{kolve2017ai2,xie2019vrgym,puig2018virtualhome,gan2020threedworld,xiang2020sapien,shen2020igibson,xia2018gibson,savva2019habitat}. However, some challenges~\cite{chakraborti2017ai} arise in the collaboration between humans and cognitive AI, as these agents may be driven by internal motivations~\cite{bach2011motivational} that are not readily apparent to humans. Consequently, many aspects of life and production, including service robots~\cite{jayawardena2010deployment}, remote training~\cite{lai2009evaluation}, interactive exhibitions~\cite{danks2007interactive}, digital assistants\cite{lu2019observational}, companion robots~\cite{gross2015robot}, and virtual communities~\cite{crang2016researching}, have been transformed by the advancement of cognitive AI.

Explainable AI (XAI)~\cite{das2020opportunities} is a burgeoning field of research within the AI community, aimed at enhancing human trust in intelligent machines. Numerous studies have focused on elucidating AI models, encompassing both local explanations~\cite{erhan2010understanding,simonyan2013deep,bach2015pixel,ribeiro2016should,lundberg2017unified} and global explanations~\cite{kim2018interpretability,lapuschkin2019unmasking,ibrahim2019global,agarwal2021neural}. 
These efforts have facilitated the development of more transparent and trustworthy AI models, thereby fostering greater human trust through more comprehensible machine behavior~\cite{rahwan2019machine}.

\subsection{Physical-Virtual Interaction}

Virtual environments facilitate interaction between physical and virtual spaces, enabling humans to access them through specific devices. For instance, a physical-virtual interaction device introduced by Brooks~\etal~\cite{brooks1990project} allowed humans to interact with virtual objects via haptics, while a directional perception approach developed by Berm{\'u}dez~\etal~\cite{bermudez2018magnetosensitive} enabled interaction with magnetic fields.

Interactive virtual environments enhance learning from demonstration~\cite{argall2009survey} by providing rich interactions between physical and virtual worlds. In the context of symmetrical reality, interaction between physical and virtual worlds is bidirectional, although few prior studies have achieved this objective. Commercial handles~\cite{lin2016virtual}, custom glove-based systems~\cite{liu2017glove}, and hybrid tracking systems~\cite{lee2021visual} can track hand poses to transmit hand information to virtual environments, representing the forward interaction of SR. Once interaction information is mapped to the virtual environment, imitation learning~\cite{zhang2018deep} becomes feasible based on observed human behavior~\cite{bates2017line}. However, the application of learned knowledge to the physical environment and the corresponding modification of physical objects remain unresolved challenges, despite the relative ease of achieving bidirectional information display~\cite{feiner1993windows,feiner1993knowledge,milgram1993applications}.

\subsection{Human-AI Interaction}

A primary objective of AI development is to serve and enhance human life, rendering the study of human-AI interaction a critical issue for AI researchers and developers. Amershi~\etal~\cite{amershi2019guidelines} proposed 18 design guidelines for human-AI interaction across diverse scenarios, providing valuable guidance for developers in designing effective interaction systems for human-AI tasks. Because AI machines are not natural entities within the physical world, they must convey information through human-comprehensible visualizations~\cite{liu2018interactive}, establishing an information bridge between humans and machines via shared workspaces~\cite{qiu2020human}. Moreover, for complex tasks, mind modeling~\cite{gao2020joint} and capability calibration~\cite{gao2022show} may be necessary to facilitate mutual understanding between humans and machines. Notice that the AI encompassed within the context of human-AI interaction includes both human-like AI (\ie, physical humanoid robots and virtual humans) and general-form AI (\ie, smart wearable devices, digital assistants, and virtual agents in the virtual world). Therefore, human-AI interaction within SR must account for these diverse AI forms to ensure a comprehensive theoretical foundation.

\begin{figure}[tb]
 \centering % avoid the use of \begin{center}...\end{center} and use \centering instead (more compact)
 \includegraphics[width=0.8\columnwidth]{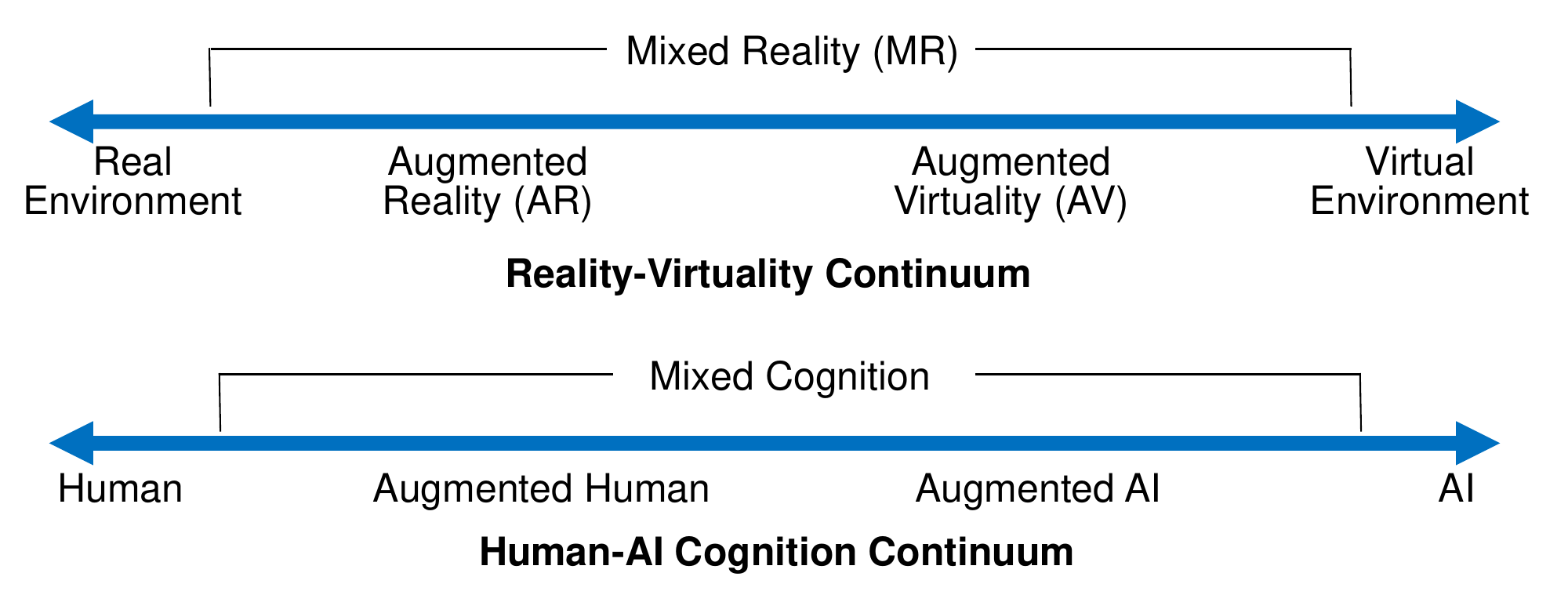}

 \caption{
 The cognition continuum. (1) The top row represents the virtuality continuum proposed by Milgram~\etal~\cite{milgram1994taxonomy,milgram1995augmented}. (2) The bottom row depicts the cognition continuum of humans and AI. Analogous to the definition of Mixed Reality, Mixed Cognition refers to the proportion of human and AI cognition within a single system.}
 \label{fig:cc}

\end{figure}

\subsection{Foundational Frameworks}

%Symmetrical reality represents a framework for modeling the phenomenon of human-AI coexistence within the context of physical and virtual worlds. 
The concept of mirror worlds~\cite{gelernter1993mirror,ricci2015mirror}, which explores the connection between physical and virtual worlds, has been discussed for many years, inspiring researchers to investigate the virtual world and yielding a series of studies on various forms of ``reality''. For example, Mann~\etal~\cite{mann2018all} surveyed numerous research works on different mixed forms (\eg, VR, AR, MR, and XR) before proposing ``All Reality'', which encompasses interactive multimedia-based ``reality'' for multiple senses. Other works, such as the ``Dual Reality''~\cite{lifton2009dual} and ``One Reality''~\cite{roo2017one} systems, have also attempted to develop specific frameworks, but have neglected the role of AI agents in disrupting human-centered paradigms by introducing new perceptual centers. Additionally, Digital Twin~\cite{boschert2016digital,haag2018digital}, extended reality (XR)~\cite{ratcliffe2021extended,stanney2021extended,coleman2009using}, and extended artificial intelligence~\cite{wienrich2021extended} differ from SR in that they do not account for AI with cognitive abilities. Further details will be provided in Subsection~\ref{sec:compare_concept}.

\section{Framework} \label{sec:framework}

\subsection{The Origin of Symmetrical Reality}

The term ``Reality-Virtuality Continuum''~\cite{milgram1994taxonomy,milgram1995augmented}, widely recognized within the VR community, describes the continuously evolving forms between real and virtual environments, as illustrated in the top row of Fig.~\ref{fig:cc}. Skarbez~\etal~\cite{skarbez2021revisiting} revisited Milgram's definition~\cite{milgram1994taxonomy,milgram1995augmented} and redefined the continuum by incorporating external virtual environments, while also addressing the discontinuity between external virtual environments and the right end of the continuum. Despite the emergence of various extensions, the dimension spanning real to virtual environments has been widely adopted by researchers.

In a similar vein, we introduce the ``Human-AI Cognition Continuum'' to represent the continuously evolving forms between human and AI cognition. Within the bottom row of Fig.~\ref{fig:cc}, ``Augmented Human''~\cite{fass2012augmented} denotes a scenario in which AI assists humans in tasks such as autonomous driving~\cite{levinson2011towards}, while ``Augmented AI''~\cite{gorban2018augmented} describes a scenario in which humans instruct AI to complete tasks such as intelligent manufacturing~\cite{hu2019irobot}. The transition between humans and AI along the cognitive axis gives rise to symmetrical reality. The proposed symmetrical reality framework extends current human-centered frameworks and benefits human society through explaining new phenomena and addressing new challenges.

\begin{figure}[tb]
 \centering % avoid the use of \begin{center}...\end{center} and use \centering instead (more compact)
 \includegraphics[width=0.8\columnwidth]{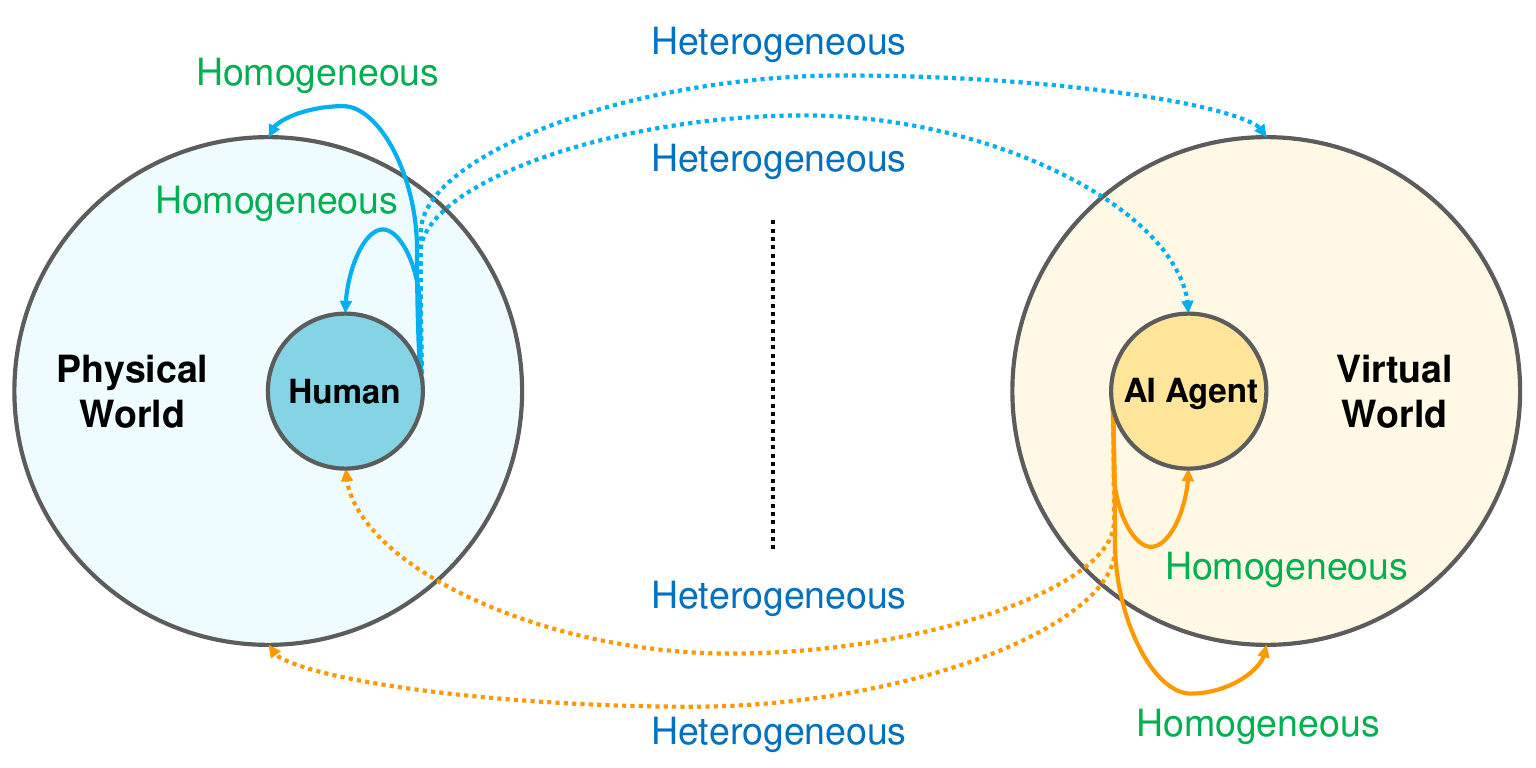}

 \caption{
 Topological symmetry of symmetrical reality. Humans, as components of the physical world, are capable of perceiving both the physical and virtual worlds, including themselves. Similarly, AI agents, as components of the virtual world, are capable of perceiving both the physical and virtual worlds, including themselves. The physical and virtual worlds are heterogeneous with respect to their forms of existence, while entities within the same world are homogeneous.}
 \label{fig:symmetry}

\end{figure}

\subsection{Definition of Symmetrical Reality}
Symmetrical reality is proposed due to its symmetrical structure regarding existing \textbf{objects} and the \textbf{interaction} between objects. The meanings of terms in the context of symmetrical reality include:

\begin{itemize}
    \item \textbf{Object}: An entity that exists in both the physical and virtual worlds, encompassing minds, embodiments, and other general objects.
    \item \textbf{Human/AI Mind}: A type of object possessing perceptual and decision-making capabilities, \ie, human minds in the physical world and AI minds in the virtual world.
    \item \textbf{Local/Remote Embodiment}: A type of object that embodies human/AI minds in local or remote spaces, facilitating perception and decision implementation.
    \item \textbf{Interaction}: Any form of information exchange of objects.
\end{itemize}

To provide an overview of the structure of symmetrical reality, as depicted in Fig.~\ref{fig:teaser}, there exist two systems with distinct perceptual centers, \ie, the human-centered system and the AI-centered system. Specifically, AR and VR fall within the domain of human-centered systems, in which the physical human serves as the perceptual center. Conversely, IAR and IVR belong to AI-centered systems, in which AI constitutes the perceptual center. Thus, topological symmetry in both perception and interaction processes underlies the essence of symmetrical reality, as illustrated in Fig.~\ref{fig:symmetry}, without necessitating similar appearances or mirrored configurations.

\begin{figure}[tb]
 \centering % avoid the use of \begin{center}...\end{center} and use \centering instead (more compact)
 \includegraphics[width=0.8\linewidth]{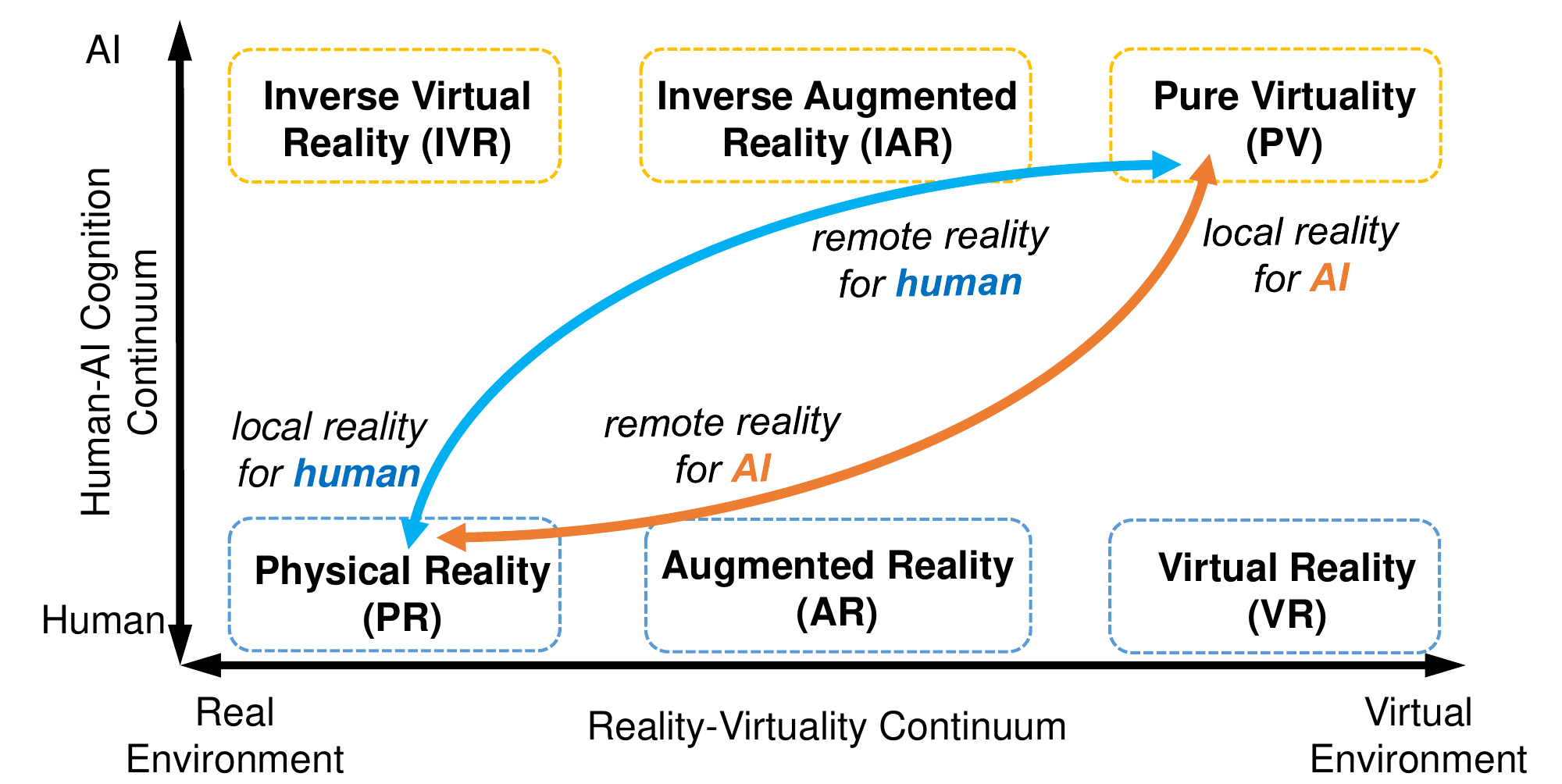}

 \caption{
 The reality framework based on the combination of the virtuality continuum and the cognition continuum.}
 \label{fig:reality}

\end{figure}

Drawing inspiration from the taxonomy of VR/AR/MR as defined by the Reality-Virtuality Continuum~\cite{milgram1994taxonomy,milgram1995augmented,milgram1999taxonomy,skarbez2021revisiting}, we propose that symmetrical reality be represented within a two-dimensional space constructed by the Reality-Virtuality Continuum and the Human-AI Cognition Continuum, as illustrated in Fig.~\ref{fig:reality}. In this figure, the physical world, also known as physical reality, constitutes the local reality for humans and the remote reality for AI. Conversely, the computer-generated virtual world, or pure virtuality, represents the local reality for AI and the remote reality for humans. The terms ``local'' and ``remote'' refer to forms of existence. In other words, physical and virtual entities are remote from one another (\ie, they are heterogeneous), while a physical entity is local to other physical entities (\ie, physical entities are homogeneous) and a virtual entity is local to other virtual entities (\ie, virtual entities are homogeneous).

Particularly, symmetrical reality emphasizes two key elements: the AI mind and its autonomous interaction. With respect to the first element, the presence of an AI mind serves as a criterion for determining the intelligence of an AI agent. According to Minsky's theory of mind~\cite{minsky1988society}, the mind may be understood as the general activity of our brains, with the intelligence of the mind arising from the vast diversity of nature. This explains why an AI agent can possess a mind through extensive learning activities, forming the foundation of the symmetrical cognitive structure of symmetrical reality. With respect to the second element, autonomous interaction serves as a criterion for determining the symmetry of information exchange. If the actions of AI are not autonomous or are fully controlled by humans, the symmetrical feature of SR cannot be sustained.
Within an SR system, information can be classified into four categories: human-generated, AI agent-generated, physical environment-generated (\eg, general vehicles without minds), and virtual environment-generated (\eg, virtual weather and natural evolution). However, environment-based information does not affect the symmetrical feature, as it is not generated by perceptual centers (\ie, humans or AI agents). In the following sections, we will introduce dual-center perception (Subsection.~\ref{sec:dual_perception}) and human-AI bidirectional interaction (Subsection.~\ref{sec:human_AI_bidirectional_interaction}), respectively.

\subsection{Dual-Center Perception} \label{sec:dual_perception}

\subsubsection{Human-Centered Perception} \label{sec:human_centered_perception}

Perceptual centers are essential for the development of minds, both in humans and AI agents. These centers allow individuals to perceive and interact with their environment. Human perception can be divided into two categories when in a symmetrical reality system.

\textbf{Human Perception of Physical Worlds.}
The study of human perception has been a popular topic for many years. When observing a purely physical world, unrelated to virtual reality, it is considered general human perception. This topic has been extensively researched in the field of cognitive science~\cite{stillings1995cognitive} and is a common occurrence in daily life.

\textbf{Human Perception of Virtual Worlds.}
Perceiving virtual worlds is more complex than perceiving physical worlds, as it requires the use of VR/AR equipment. For instance, VR head-mounted displays~\cite{jensen2018review} serve as visualization tools for virtual elements, enabling humans to perceive virtual worlds. The rules governing virtual worlds may differ from those of physical worlds, such as variations in perceived causality~\cite{wang2018spatially}. As a result, humans must acquire new knowledge through the perception of virtual worlds.

\subsubsection{AI-Centered Perception} \label{sec:AI_centered_perception}
An AI agent requires a mind~\cite{minsky1988society}. Typically, researchers design the virtual world and AI agent with a passive configuration, capable only of responding to actions within the physical world without possessing a mind. In contrast, the symmetrical reality framework regards the AI agent as an autonomous entity with equivalent status to physical humans, so AI agents become the perceptual center in the virtual world, which is called AI-centered perception.

\textbf{AI Perception of Physical Worlds.} 
The perception of physical worlds pertains to the recognition of ``\textit{what is virtual}'' and ``\textit{what is real}'' \cite{brooks1999s}. Generally, when we refer to an object as virtual, we mean that it is a computer-generated graphic or audio, distinct from physical entities. Within the context of symmetrical reality, the distinction between physical and virtual objects is defined from the perspective of humans for terminological convenience, although the AI-centered definition of what is physical or virtual should be opposite to that of humans.

\textbf{AI Perception of Virtual Worlds.}
As previously mentioned, an AI agent resides within the virtual world and thus perceives it as its local reality. This implies that AI agents and virtual objects are both situated within the virtual world and are therefore homogeneous. Possessing a mind enables the AI agent to perceive the virtual world in the same manner as a physical human perceives the physical world, facilitating improved human-AI communication. At the same time, other modes of perception, such as direct data query, may also be employed if human comprehension is not a consideration.

\subsection{Human-AI Bidirectional Interaction} \label{sec:human_AI_bidirectional_interaction}

Human-AI interaction has recently emerged as an important and promising field~\cite{rauter2019robot}. Within the framework of symmetrical reality, physical humans and AI agents can interact directly with one another, unhindered by the physical-virtual boundary. Physical humans can enter the virtual world through AR/VR, while AI agents can enter the physical world through IAR/IVR, thereby facilitating bidirectional interaction between the two perceptual centers.

\textbf{Embodiment of Humans in Virtual World.}
VR and AR technologies enable humans to enter a purely virtual or a physical-virtual mixed environment. In these scenarios, a virtual avatar is created to transmit human actions to the virtual world, thereby altering the states of virtual objects. The virtual avatar and virtual objects coexist within the same virtual space, allowing them to interact seamlessly according to the rules of the virtual world. This avatar is referred to as the embodiment of humans in the virtual world.

\textbf{Embodiment of AI in Physical World.}
Enabling AI agents to enter the physical world and interact with physical objects requires complex work. If an AI agent intends to interact with a physical object (\eg, moving a chair or grasping a cup), it must be capable of controlling a physical component (\eg, a mechanical arm) to execute these actions on its behalf. This physical component constitutes a physical embodiment of the AI agent, implementing actions determined by the agent. Typically, the physical component can take the form of a physical robot, such as a humanoid robot~\cite{ishiguro2001robovie}, industrial robot~\cite{wallen2008history}, or robot arm~\cite{bellicoso2015design}, \etc.

\begin{figure}[tb]
 \centering % avoid the use of \begin{center}...\end{center} and use \centering instead (more compact)
 \includegraphics[width=\columnwidth]{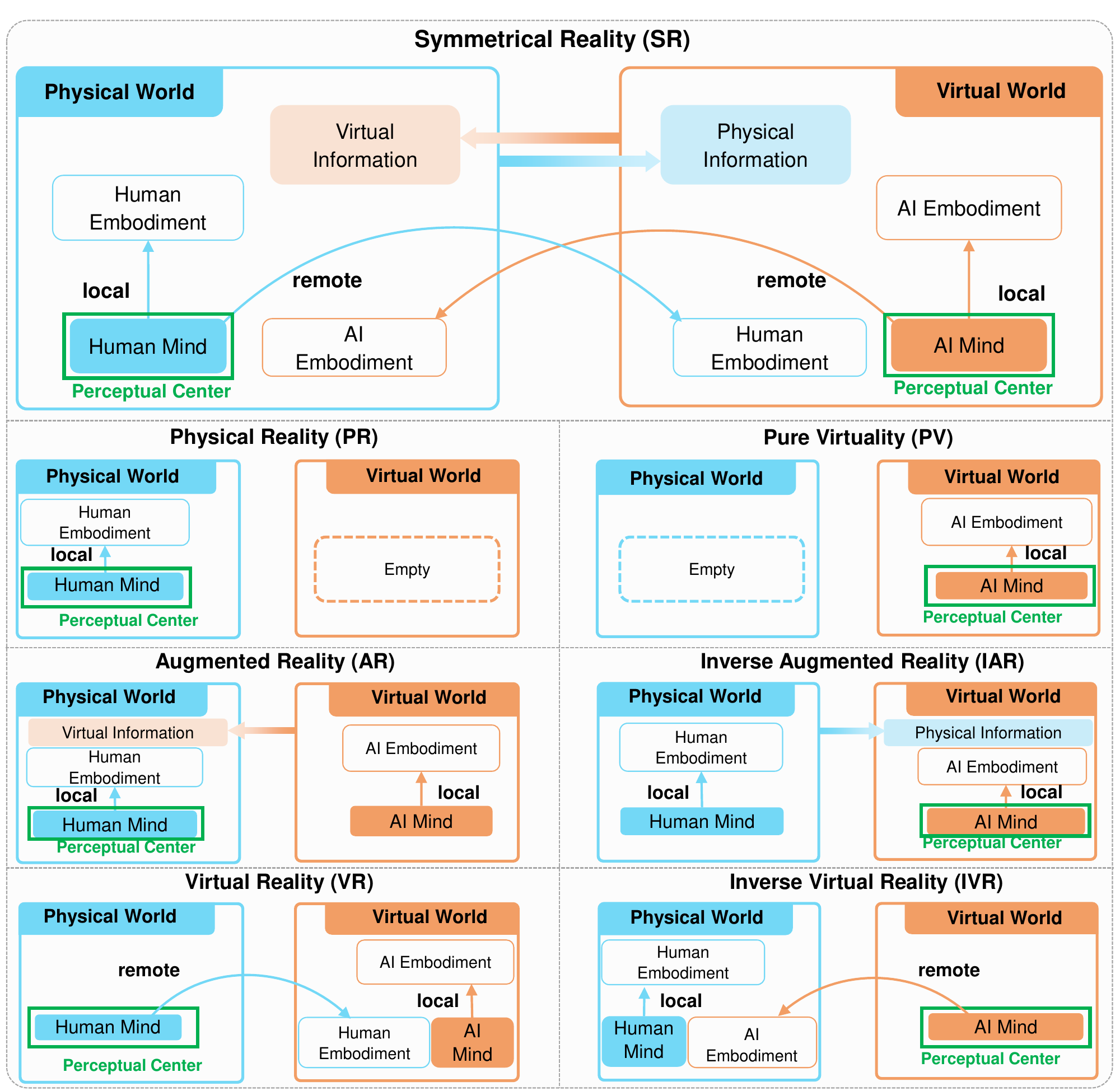}

 \caption{
 Symmetrical reality and other special forms. The top figure illustrates symmetrical reality, while the smaller figures at the bottom depict six distinct special cases, \ie, physical reality (PR), augmented reality (AR), virtual reality (VR), pure virtuality (PV), inverse augmented reality (IAR), and inverse virtual reality (IVR). According to the Reality-Virtuality Continuum proposed by Milgram~\etal~\cite{milgram1994taxonomy, milgram1995augmented, milgram1999taxonomy}, mixed reality (MR) comprises AR and Augmented Virtuality (AV). For the sake of convenience, this paper will use AR as a representative of all physical-virtual mixed frameworks.
 }
 \label{fig:virtual_real} 

\end{figure}

\subsection{Symmetrical Reality and Special Cases} \label{sec:core_components}

Fig.~\ref{fig:virtual_real} displays the core components of symmetrical reality in the top row, with six special forms that can be unified by SR shown below. When certain components are removed or disregarded, SR degenerates into less complex forms. Although SR can unify many forms of physical-virtual mixed systems, there are three specific forms that warrant discussion due to their status as collection concepts containing a variety of content. These concepts are introduced in Subsection~\ref{sec:compare_concept}.

\subsection{Comparison with Other Concepts} \label{sec:compare_concept}

Symmetrical reality is compared with three other concepts, \ie, Digital Twin, extended reality, and extended artificial intelligence.

\textbf{Digital Twin.}
The concept of ``Digital Twin''~\cite{boschert2016digital,haag2018digital} was proposed several years ago and bears some similarity to symmetrical reality. However, there are fundamental differences between the two concepts. Digital Twin is a framework that maps the physical world to the virtual world throughout its lifespan, enabling digital information to assist humans in managing their physical environments. This has led to its widespread use in industrial areas~\cite{tao2018digital}. The first difference between Digital Twin and symmetrical reality concerns existing objects. In symmetrical reality, virtual objects may exist without being mapped from the physical world and vice versa, whereas Digital Twin typically requires a complete mapping structure between the physical and virtual worlds. The second difference is the dual-center perceptual structure. In symmetrical reality, physical humans and AI agents have equal status regarding their cognitive activities~\cite{sperber1986relevance}, while Digital Twin does not address cognitive issues. The third difference concerns the system operation mechanism. In symmetrical reality, it is assumed that AI agents may possess a mind (and sometimes even free will \cite{libet1999we}), allowing them to perceive, cognize, make decisions, and act independently without human intervention. This validates the strict mapping rule between the virtual and physical worlds required by Digital Twin.

\textbf{Extended Reality.}
Extended reality (XR) \cite{ratcliffe2021extended,stanney2021extended,coleman2009using} encompasses all computer-generated environments and human-machine interactions that combine physical and virtual elements. In some literature, XR is an umbrella term that includes virtual reality, augmented reality, mixed reality, and other forms of physical-virtual integration. This technology has been applied in manufacturing \cite{fast2018testing,doolani2020review} and medicine \cite{andrews2019extended}. Since XR is designed for humans, it represents a partial description of SR from the human perspective.

\textbf{Extended Artificial Intelligence.}
Extended artificial intelligence \cite{wienrich2021extended} examines human-AI interaction based on the XR-AI continuum. The concept focuses on the combination of XR and AI, thus making XR become more intelligent. This combination serves as a new way to study the effects of prospective human-AI interaction. However, extended artificial intelligence does not address the perception and interaction of AI agents from the agent's perspective, particularly when the AI agent possesses autonomous perception and interaction capabilities. This difference in focus distinguishes extended artificial intelligence from symmetrical reality.

\section{Technical Challenges} \label{sec:tcsr}

%\subsection{General Introduction}
Symmetrical reality is a framework that describes the coexistence of humans and AI agents, as well as the bidirectional interaction between them. Although high-level AI with minds has not yet been fully realized given the current state of artificial intelligence, and even low-level AI such as digital assistants is not fully developed, it is still possible to conduct preliminary studies using simulation platforms with reasonable assumptions.

Constructing symmetrical reality systems presents significant technical challenges due to the ongoing development of many key techniques. These challenges primarily arise from the generation of AI minds, symmetrical perception, symmetrical interaction, and the avoidance of conflicts between physical and virtual signals.

\subsection{AI Minds} \label{sec:agent_sr}

Creating a mind \cite{kurzweil2013create} has long been a formidable challenge, and the generation of AI minds is the most fundamental challenge in building a symmetrical reality system. To achieve human-compatible AI, it is necessary to consider human characteristics, which may be useful in enabling AI agents to learn from humans and think and act accordingly. It is important to note that human-like AI is just one possible direction for the future and not the only one. The term ``human-like'' refers to the ability of AI agents to perceive, think, and act, rather than limiting them to the same methods as humans. Human-like AI is a typical form of AI agent and can be transformed into other general forms by adjusting structures and parameters.

\textbf{Human Perception.}
In a symmetrical reality system, human perception is based on a needs theory~\cite{maslow1943theory}. Physical humans are inherently capable of adapting to the physical world, using their eyes, ears, and hands to obtain visual, auditory, and tactile information. Similarly, in symmetrical reality, humans can use a perception module to obtain information from the virtual world and communicate with AI agents located there. The technical challenge in human perception lies in designing the user interface to account for all sensation channels and efficiently communicate with AI agents.

\textbf{Generation of AI Mind.}
The generation of AI minds must consider both human-like and general-form AI. Human-like AI should be created based on the characteristics of real humans, while general-form AI can be developed according to specific goals. Regardless of the type of AI agent, it must be able to perceive its surroundings through visual, auditory, and tactile channels in order to understand its environment. It is also essential for AI agents to possess minds that enable them to perceive both physical and virtual worlds, process information, and take appropriate actions. Building an AI mind is challenging due to our incomplete understanding of how the human brain works and the limitations of state-of-the-art AI models \cite{bommasani2021opportunities,brown2020language} in performing complex tasks. Even ChatGPT \cite{zhuo2023exploring} and GPT-4 \cite{openai2023gpt4}, despite their impressive performance in language tasks, do not fully solve the problem of mind simulation.

\subsection{Symmetrical Perception}
Unlike VR and AR systems, a symmetrical reality system is designed to have two perceptual centers, one located in the physical world and the other in the virtual world. The physical human, who exists in the physical world, serves as one perceptual center, while the AI agent, which exists in the virtual world, serves as the other. This structure enables forward perception by humans (corresponding to human-centered perception in Subsection~\ref{sec:dual_perception}) and inverse perception by AI agents (corresponding to AI-centered perception in Subsection~\ref{sec:dual_perception}), creating symmetrical perception between humans and AI.

\textbf{Forward Perception.}
Forward perception refers to the traditional perceptual process initiated by humans. Specifically, when a human is situated within a symmetrical reality system, he/she can observe both physical and virtual elements within the system. This would be a typical VR or AR system if nothing else were considered. However, it is important not to overlook the presence of the AI agent, as it possesses a mind that enables it to act autonomously.

\textbf{Inverse Perception.}
Inverse perception is the reverse of forward perception and describes the perception of AI agents in the virtual world. In a symmetrical reality system, an AI agent can perceive both its virtual world and the physical world, as described in Subsection~\ref{sec:core_components} as IAR or IVR.

The technical challenges arise in two ways. Human forward perception may be affected by AI agents if the virtual world changes unexpectedly. Conversely, the inverse perception of AI agents may be constrained by humans due to ethical and privacy concerns. However, minimizing inverse perception also reduces the capabilities of AI agents and diminishes the services they can provide. Balancing these gains and losses presents a challenge for the community.

\subsection{Symmetrical Interaction}

Interaction between the physical and virtual worlds is achieved by processing signals from two sources. Signals from the physical world are generated by physical humans and sent to either the physical or virtual world, while signals from the virtual world are generated by AI agents and sent to either the virtual or physical world, creating symmetrical interaction. Subsection~\ref{sec:human_AI_bidirectional_interaction} has shown the details of human-AI bidirectional interaction, while this subsection discusses the challenges of symmetrical interaction.

\textbf{Signals from Physical World.}
Humans generate signals for two purposes: to interact with the physical world and to interact with the virtual world, including AI agents. For example, humans may need to obtain food and water in the physical world, or they may use VR devices to enter the virtual world and manipulate virtual objects, such as moving a virtual chair.

\textbf{Signals from Virtual World.}
Like physical signals, AI agents can generate signals to interact with other parts of the virtual world and the physical world. When an AI agent manipulates virtual objects, the action can be performed directly because both the agent and object belong to the virtual world. However, when an AI agent seeks to manipulate physical objects, it must use physical components as a bridge to effect changes in the physical world.

The challenges of symmetrical interaction arise primarily from mutual control between the physical and virtual worlds. Physical-to-virtual control can be achieved by programming physical changes into virtual spaces, while virtual-to-physical control requires mechanical systems or other modules capable of affecting physical environments. Integrating bidirectional control presents significant difficulties in both design and implementation.

\subsection{Addressing Conflicts}
When a system contains multiple decision-making units, conflicts may arise. In the context of symmetrical reality, cooperation between different agents and the halting problem are critical issues that must be addressed.

Symmetrical reality features two perceptual centers and supports bidirectional interaction. This dual-center structure may encounter decision conflicts when the physical human and AI agent take different actions regarding the same object. For example, consider a mug on a table in the physical world and its corresponding virtual mug on a virtual table. The physical and virtual mugs can be strongly mapped to each other, allowing humans to move the physical mug and AI agents to move the virtual mug. If the human and AI agent simultaneously move the mug, a conflict will occur and the system may crash. Therefore, tasks in symmetrical reality systems must account for cooperation and conflicts to ensure normal operation.

Given the equivalent status of physical humans and AI agents in symmetrical reality systems, it is essential to find effective solutions for managing conflicts.

\begin{figure}[tb]
 \centering % avoid the use of \begin{center}...\end{center} and use \centering instead (more compact)
 \includegraphics[width=0.9\columnwidth]{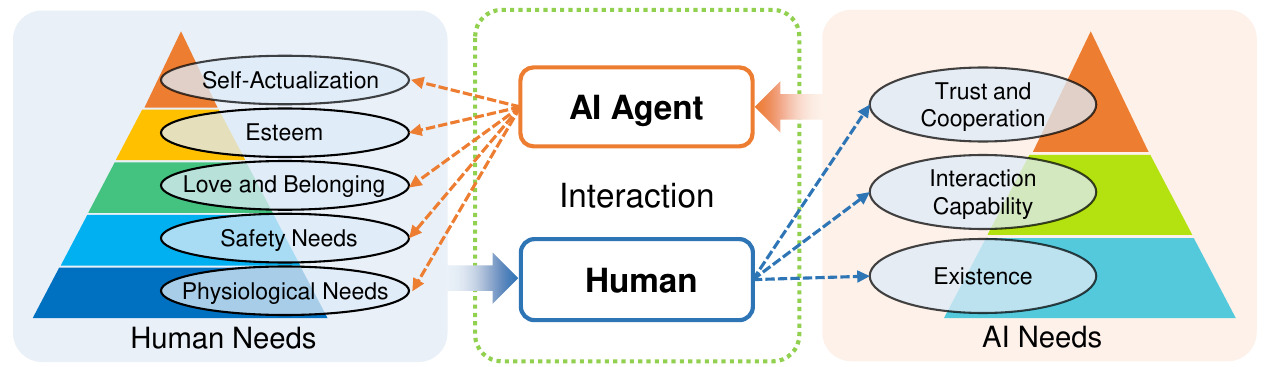}

 \caption{
 Human-AI coexistence in symmetrical reality. The AI agent can influence the human's long-term living paradigm on five levels, enhancing the human's overall experience. Conversely, humans can also affect the AI agent on three levels.}
 \label{fig:need}

\end{figure}

\section{Coexistence of Humans and AI} \label{sec:coexistence}

Symmetrical reality enables long-term coexistence between humans and AI agents, with human-compatible AI necessary to ensure the safety and comfort of human users.

\subsection{Human-AI Coexistence}

Long-term coexistence entails humans and AI agents living together within a symmetrical reality system, with their needs satisfied. This means that a human is immersed in a virtual or physical-virtual mixed environment for an extended period via wearable devices \cite{guo2019evaluation}, while an AI agent exists in the same environment for the duration. The general long-term setting typically involves several hours of continuous VR/AR device use \cite{guo2020exploring}, necessitating the provision of basic needs to ensure normal living for humans and normal functioning for AI agents.

During long-term coexistence, humans have many basic needs, such as eating, drinking, and sleeping. To analyze these needs, Maslow's hierarchy of needs theory \cite{maslow1943theory} is applied. Researchers have studied the challenges of long-term living in virtual reality, providing insights into potential issues and new problems that may arise during human-AI coexistence. Intuitively, designing a long-term coexistence system requires satisfying human needs from lower to higher levels, as well as addressing the diverse needs of AI if needed. Based on this design, physical humans and AI agents can coexist and interact with each other.

\subsection{Needs Theory in Symmetrical Reality}

%AI agents with the ability to think and act may influence human behavior. 
Human and AI needs differ due to their distinct forms of existence. A detailed discussion of the rationality of human and AI needs is beyond the scope of symmetrical reality, so we adopt Maslow's hierarchy of needs for humans and make reasonable adaptations for AI agents. Human needs are divided into five levels, while AI needs are divided into three levels, as shown in Fig.~\ref{fig:need}. The needs theory for AI agents can be very diverse, which will be discussed in detail later. The interaction and mutual influence between humans and AI agents prompt a reconsideration of the human-AI relationship~\cite{zhang2020machine}, particularly within the framework of symmetrical reality.

\subsubsection{Human Needs}
According to Maslow's hierarchy of needs theory, human needs can be classified into five levels as follows.

\textbf{Physiological Needs.}
Even when immersed in a symmetrical reality system, humans still have physiological needs. A person must eat and drink, even in a virtual or physical-virtual mixed environment. The symmetrical reality system (\eg, a head-mounted display-based system) should support basic human needs such as food and drink.

\textbf{Safety Needs.}
The prospect of human-AI long-term coexistence may give rise to safety considerations. Symmetrical reality, by mapping segments of the virtual world onto corresponding areas of the physical world, facilitates rapid human adaptation to the virtual environment. This helps address concerns related to physical safety and psychological health.

\textbf{Love and Belonging.}
Emotional needs are essential for humans, and long-term coexistence requires providing an emotional experience for users. The independent AI agent in a symmetrical reality system can serve as a friend and companion to the human user. Family is important to humans, providing love and support. Being situated in a virtual environment without other people might be emotionally harmful. This is where symmetrical reality can be useful. AI agents with minds can be embodied as virtual humans, serving as friends or family to users in the virtual world.

\textbf{Esteem.}
Humans require esteem, even when living in a virtual environment. Esteem involves receiving respect from others and having the freedom to engage in regular activities. In daily life, humans seek esteem from others to facilitate cooperation. This social need can be met by AI agents that adopt human appearances and behavior patterns.

\textbf{Self-Actualization.}
Humans in virtual environments also need self-actualization, with the desire for excellence enhancing the immersive experience. Cooperation between physical humans and AI agents can benefit both parties. For example, an AI agent may actively seek out interesting or useful information, attracting human attention to cooperative tasks and increasing immersion.

\subsubsection{AI Needs}
Similar to the definition of human needs, we define the goals or objectives pursued by AI as ``AI needs''. In the interest of simplicity, we focus on essential AI needs in human-AI interaction.

It is important to note that AI can exist in various forms, resulting in dramatically different AI needs. Some general-form AI may have no needs at all, depending on their roles in daily life. Nonetheless, humans can design AI structures to imbue AI with certain ``needs''. Therefore, we define three levels of AI needs for symmetrical reality: existence, interaction capability, and trust and cooperation. This definition is not exhaustive, as AI may have entirely different needs or none at all. However, it represents an initial hypothesis of potential AI needs, which should be useful for cooperative human-compatible AI agents.
Drawing on Maslow's hierarchy of needs theory for humans, we propose a possible way to define AI needs by mapping the first and second levels of human needs (physiological and safety needs) to the AI need as ``existence'', the third level of human needs (love and belonging) to the AI need as ``interaction capability'', and the fourth and fifth levels of human needs (esteem and self-actualization) to the AI need as ``trust and cooperation''. Of course, there may be various ways to define AI needs, including definitions unrelated to human needs due to the specific forms of AI.

\textbf{Existence.}
Existence is the most fundamental need for an AI agent, without which it cannot function. This is analogous to the first two levels of human needs, which ensure survival and safety. Intuitively, an AI agent must take every possible action to maintain its existence in its environment, unless such actions violate AI ethics or other fundamental principles.

\textbf{Interaction Capability.}
AI agents must be able to perceive and interact with the physical and virtual worlds to understand and affect their external environments. The ability to interact with external environments is essential for establishing connections with humans and distinguishes AI agents from ordinary virtual objects.

\textbf{Trust and Cooperation.}
Acquiring knowledge helps AI agents understand humans and gain their trust, while acquiring skills facilitates smoother human-AI cooperation. Knowledge can be transferred from humans to AI agents through demonstration, such as showing an AI agent how to cook or clean. This knowledge concerns the abstract logic of task implementation, allowing knowledge generated by physical humans to be directly transferred to the virtual world. Skills are well-trained actions based on knowledge and are acquired through repeated observation of human behavior or self-exploration within the environment. 
%Both humans and AI agents are capable of manipulating physical and virtual objects. This equivalent interaction facilitates information exchange across different spaces and enables smooth cooperation between AI agents and humans.

\begin{figure}[tb]
 \centering % avoid the use of \begin{center}...\end{center} and use \centering instead (more compact)
 \includegraphics[width=\linewidth]{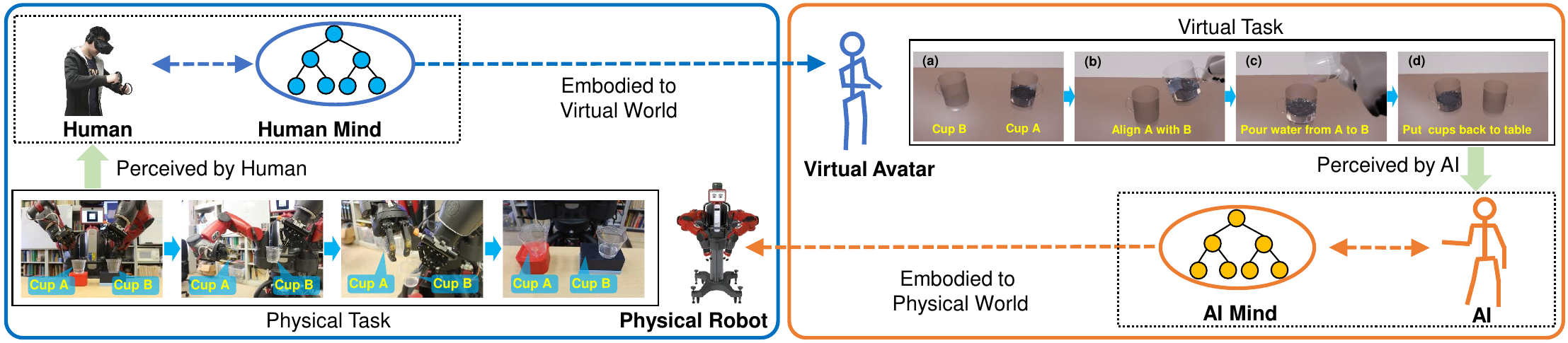}

 \caption{
 Symmetrical reality system with a simulated AI mind and bidirectional interaction. A human can demonstrate a task, such as pouring water, to an AI agent within the virtual world. The AI agent can then learn how to perform this task and apply the knowledge in the physical world.} 
 \label{fig:SR_sim}

\end{figure}

\subsection{Conceptual Implementation of Symmetrical Reality} \label{sec:sr_system}

Two operations are necessary to demonstrate the functioning of symmetrical reality. The first is to develop a model that simulates the AI agent's mind, enabling it to think and act autonomously, even at a relatively low level. The second is to implement bidirectional interaction between the physical and virtual worlds. 
Pouring water \cite{pan2017feedback} is a typical daily task that illustrates the interaction between humans and AI agents across physical and virtual worlds. We use this task to explain how a symmetrical reality system is constructed and operates. As shown in Fig.~\ref{fig:SR_sim}, we employ popular graphical models \cite{koller2009probabilistic} to represent task knowledge, with the mind also visualized as a graphical structure. Driven by its mind, the AI agent has the autonomy to decide whether, when, and how to pour a cup of water.

\subsubsection{AI Mind Simulation}

A graph-based approach is adopted to simulate the AI mind for the task of pouring water, which provides the foundation for bidirectional interaction between humans and AI agents. We employ a spatial, temporal, and causal And-Or Graph (STC-AOG)~\cite{zhang2020graph} to represent the internal workflow of the AI mind during task performance. While this is not the only possible method for this task (\eg, deep neural networks~\cite{lecun2015deep} may also be applicable), it serves as a useful example due to its interpretability.

\subsubsection{Bidirectional Interaction} \label{sec:bidirectional_interaction}

Bidirectional interaction can be achieved within a symmetrical reality system, as shown in Fig.~\ref{fig:SR_sim}. A human can perform the task of pouring water, with their virtual avatar simultaneously controlling the virtual cup and water in the virtual world. Similarly, an AI agent can perform the task of pouring water in the virtual world, with its physical embodiment (\ie, a physical robot) performing the same action in the physical world. This process demonstrates the bidirectional interaction of symmetrical reality, with the AI agent driven by a simulated mind, showing that both humans and AI can perform tasks across physical and virtual worlds. 
%For further details beyond this brief introduction, including mind simulation and bidirectional interaction, please refer to the supplementary materials.

\section{Proof-of-Concept Scenarios} \label{sec:poc}
\subsection{Exemplary System Designs}
Symmetrical reality is a new concept that describes the coexistence of humans and AI across physical and virtual worlds. The complexity of the system can make implementation less intuitive. 
Inverse virtual reality and inverse augmented reality are two special cases of symmetrical reality that serve as useful examples for understanding the design of symmetrical reality systems.

\begin{figure}[tb]
 \centering % avoid the use of \begin{center}...\end{center} and use \centering instead (more compact)
 \includegraphics[width=\columnwidth]{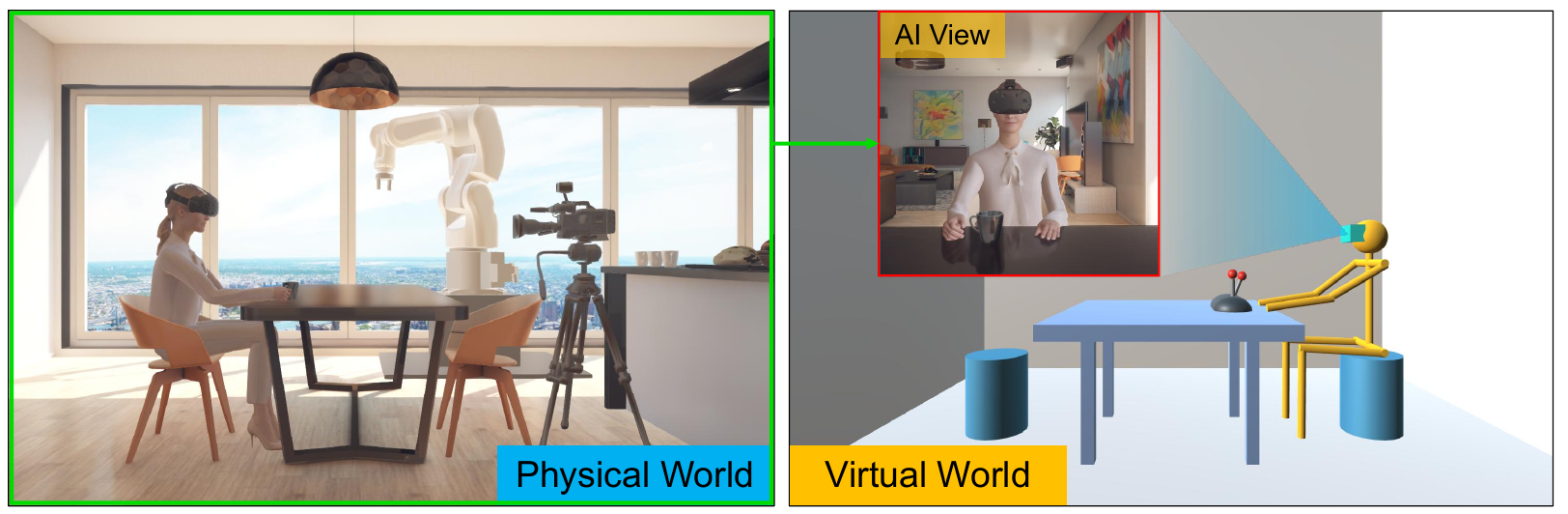} 

 \caption{
  Overview of the inverse virtual reality system. The left figure depicts the physical world, while the right figure depicts the virtual world. The AI agent is immersed in the physical world and can manipulate physical objects using a robot arm.}
 \label{fig:ivr}

\end{figure}

\textbf{Inverse Virtual Reality.} 
Virtual reality systems have been around for many years, with some designed to provide high levels of immersion for human users. Analogously, we have designed an inverse virtual reality system for AI agents. The system configuration is shown in Fig.~\ref{fig:ivr}. In the physical world, depicted on the left side of the figure, a human sits at a table wearing a head-mounted display that shows virtual information. A camera opposite the human captures visual information from the physical space and sends it to the AI agent. A robot arm is set up to assist the AI agent in manipulating physical objects, enabling virtual-to-physical interaction by the AI agent. In the virtual world, depicted on the right side of Fig.~\ref{fig:ivr}, an AI agent observes physical information captured by the camera, with its view filled exclusively with physical content. The AI agent can act in the physical world by controlling the robot arm, such as using it to pick up a cup of coffee and place it in front of the human, providing a service that spans physical and virtual worlds.

\begin{figure}[tb]
 \centering % avoid the use of \begin{center}...\end{center} and use \centering instead (more compact)
 \includegraphics[width=\columnwidth]{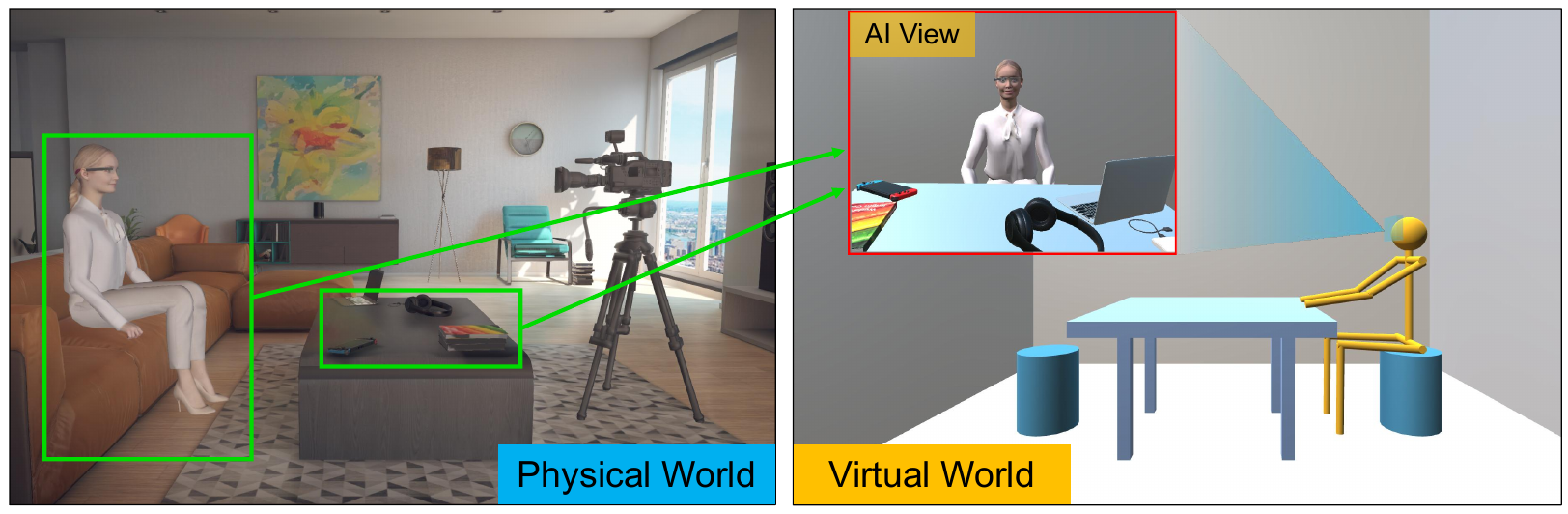}
 
 \caption{
 Overview of the inverse augmented reality system. In the left figure, the human and objects outlined in green are mapped to the virtual world shown in the right figure. The AI agent can see both the virtual environment and physical objects.
 }
 \label{fig:iar}

\end{figure}

\textbf{Inverse Augmented Reality.}
Augmented reality, which focuses on creating physical-virtual mixed environments for humans, has been around for years. However, inverse augmented reality, which creates physical-virtual mixed environments for AI agents, has not been formally discussed. We have designed a typical scenario to demonstrate how inverse augmented reality develops and operates, as shown in Fig.~\ref{fig:iar}. In this experimental setup, a human sits on a couch with personal items such as a book and a laptop on a nearby tea table. A camera captures images of the human and personal items, and then maps them to the virtual world. In the virtual world, the AI agent can see not only mapped objects from the physical world (\eg, the human and personal items) but also virtual objects (\eg, a virtual table and walls) in its own virtual world. As a result, the AI agent's view is a mixed scene of physical and virtual elements created by augmenting physical objects into virtual space.

\subsection{Potential Applications}
\begin{figure}[tb]
    \centering
    \includegraphics[width=\linewidth]{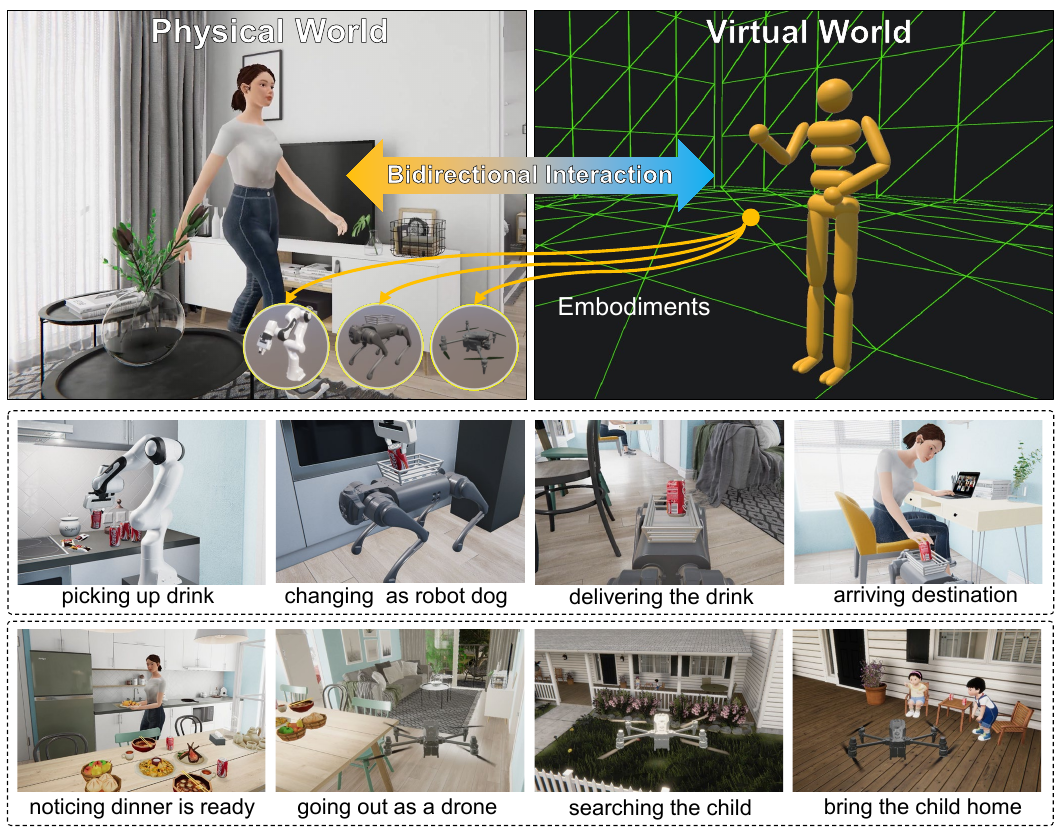}

    \caption{Illustration of the potential applications. The figure at the top provides a schematic representation of the system’s architecture. The middle and bottom figures serve as practical demonstrations of the system’s capabilities, specifically in the context of beverage service and child location. This application example illustrates that the AI agent is capable of assimilating human behavioral patterns through daily observation, subsequently offering reasonable assistance as per human requirements. This holds true irrespective of whether the need arises in the physical or virtual world.}

    \label{fig:application}
\end{figure}

To show the potential applications of symmetrical reality, we propose some scenarios that could occur in everyday human life. Given the intuitive nature of how humans enter into the virtual world, our demonstration primarily highlights the mechanisms by which AI agents can transition into the physical world through a variety of embodiments. As depicted in Fig.~\ref{fig:application}, the AI agent can coexist with humans and proactively offer assistance when there is a possibility that the human may require help. The proposed application scenario primarily underscores the daily assistance (such as beverage service and child location) provided by the AI agent, which enters the physical world through diverse embodiments (\eg, robot arms, robot dogs, and drones). Certainly, it also accommodates the possibility of humans transitioning into the virtual world when necessary.

In addition to the general scenarios mentioned above, certain specialized applications, including smart healthcare~\cite{tian2019smart} and smart education~\cite{zhu2016research}, can leverage the bidirectional framework of symmetrical reality that spans both physical and virtual worlds. In such instances, AI agents have the potential to deliver prompt and expert services via both physical and virtual embodiments~\cite{sun2023neighbor}. For example, AI agents can not only interact with humans in the virtual world through the generation of graphical or auditory signals, but also execute physical actions by controlling robots or other mechanical devices to offer tangible services to human society, such as managing emergencies. This form of autonomous assistance introduces a novel approach to addressing unforeseen circumstances that frequently arise in human society.

\section{Discussion and Conclusion} \label{sec:discussion}

\subsection{Future Directions}

The emergence of symmetrical reality is driven by the symmetrical existence and interaction of objects, necessitating the autonomy of AI agents in the virtual world. In practice, when constructing a symmetrical reality system, two types of AI agents typically exist in the virtual environment~\cite{mateus2016intelligent}: character agents (\eg, human-like AI) and environment agents (\eg, general-form AI). Character agents can learn from human behavior \cite{hattori2010learning} through learning paradigms such as evolutionary computation \cite{fogel2006evolutionary} and reinforcement learning \cite{sutton2018reinforcement}, enabling direct human-AI interaction. Environment agents typically operate according to predefined configurations and perform specific tasks. It is important to note that the boundary between environment and character agents is not always clear, and they can be considered different embodiments of AI agents.

The importance of studying symmetrical reality can be understood in terms of the level of artificial intelligence. While high-level artificial intelligence, particularly artificial general intelligence (AGI) \cite{goertzel2014artificial}, has not yet been fully realized, this has not diminished society's interest in AI development and related research. Many studies are conducted before real situations arise, providing valuable information in advance and prepare appropriate solutions. By building experimental systems, advanced research topics related to AGI and human-AI coexistence can be explored.

%The proof-of-concept designs presented in Section~\ref{sec:poc} provide researchers and developers with a reference for constructing symmetrical reality systems. While the framework is relatively complex, independent modules can be developed and integrated to build prototypes for various tasks. Implementation will depend on the feasibility of specific designs and the robustness of engineering.

Symmetrical reality is not limited to the context of XR and can involve interaction with AI through various forms of displays. Whether humans are immersed in a virtual world or AI is embodied in the physical world does not alter the structure of symmetrical reality, as the existence and inherent attributes of humans and AI already validate the SR framework. The XR example presented in this paper is just one of many possible system forms. It is also important to consider real-world scenarios, such as personal computers, smartphones, smart homes, and other devices related to the internet of things \cite{li2015internet} and ubiquitous computing \cite{krumm2018ubiquitous}. These common real-world scenarios also serve as special cases of symmetrical reality, as the devices within them connect the physical and virtual worlds, enabling potential bidirectional perception and interaction.

\subsection{Open Problems}

Symmetrical reality emphasizes the symmetry between humans and AI agents in terms of perception and interaction, representing a form of topological symmetry. No other symmetry is essential, and asymmetries can exist in many aspects. For example, not every physical object must have a corresponding virtual object, and events may occur exclusively in the virtual world without any relation to the physical world. These asymmetries within symmetrical reality increase the functional diversity of the physical and virtual worlds. If the essential symmetry is broken, such as by the absence of an AI agent, the symmetrical reality system will degenerate into one of its special forms. If the virtual world is entirely new and unrelated to the physical world, it represents a maximization of asymmetry within symmetrical reality, which is an extreme case. %Symmetrical reality aims to encompass all physical-virtual settings by adjusting its parameters.

Cognitive AI is a typical presence in a symmetrical reality system, corresponding to a certain level of human-like AI. As a complement, non-cognitive AI should also be compatible with these situations, corresponding to general-form AI. Non-cognitive AI may behave differently from humans and may not have any meaningful embodiment, but it can perform specific tasks well. For example, a recommendation system can make many decisions to enhance human experiences in various applications without requiring human-like behavior or embodiment. Symmetrical reality systems should be able to leverage this dedicated AI to improve their service capabilities.

Human-level AI is a goal of the AI community, and the proposed symmetrical reality framework also anticipates such an AI agent for cooperation with humans. With the development of increasingly advanced models and algorithms \cite{zhuo2023exploring,openai2023gpt4}, the capabilities of AI agents are rapidly increasing. Regardless of whether human-level or other forms of strong AI are ultimately achieved, society has observed a trend of increasing AI capability. If an AI agent is at a relatively low level, the performance and user experience of symmetrical reality will be correspondingly diminished, but its main features can still be maintained to some extent. If AI develops into a high level, high-level symmetrical reality will also be realized, providing a corresponding framework for describing human-AI coexistence.

The ultimate goal of AI agents should not be limited to emulating humans, as humans have inherent physical limitations based on their biological structure. 
For example, the perception capabilities of AI should not be confined to human senses. AI can be engineered as a potent entity with respect to computation, storage, evolution, perception, and reasoning.
Symmetrical reality aims to encompass different types of AI, including both human-like and general-form rational machines. Since humans are unique and typical intelligent agents, we primarily use human-like AI agents to introduce the SR framework, but it is important to note that all types and levels of intelligence can interact with humans. Regardless of its type or level of intelligence, SR should be able to accommodate human-AI interaction. In terms of limitations of the SR framework, if AI is at a low level of intelligence, its ability will be limited and SR features (\eg, dual-center perception) may degenerate.

Given the pivotal role of AI agents in symmetrical reality, it is important to prioritize AI fairness~\cite{mehrabi2021survey} to prevent the emergence of bias. To tackle this issue, it is necessary to establish a standard test~\cite{peng2023tong} for evaluating value-driven AI systems in intricate environments, thereby fostering value alignment~\cite{yuan2022situ} between humans and AI agents. Reliability~\cite{ryan2020ai} presents another challenge. Symmetrical reality intertwines humans and AI agents in both physical and virtual worlds, necessitating a high degree of AI reliability to ensure safe human-AI interaction.

\subsection{Conclusion} \label{sec:conclusion}

This paper introduces the emergence of symmetrical reality and presents a comprehensive framework by analyzing its essential components, with dual-center perception and bidirectional interaction identified as key features that demonstrate the value of symmetrical reality. Technical challenges in developing symmetrical reality systems are discussed, followed by an analysis of human-AI coexistence within symmetrical reality. Demonstrations of symmetrical reality systems and proof-of-concept scenarios provide insights for researchers on the practical application of symmetrical reality.

While symmetrical reality is still in its nascent stages and confronts numerous challenges, it is a field that deserves increased focus. Tackling the myriad challenges inherent in symmetrical reality will necessitate interdisciplinary cooperation across various domains, including mathematics, physics, neuroscience, psychology, computer science, electronic engineering, and mechanical engineering.

% \begin{ack}
% This work was supported by the National Natural Science Foundation of China (No. 52305007, No.61976214).
% \end{ack}

\bibliographystyle{abbrv}
%\bibliographystyle{abbrv-doi-narrow}
%\bibliographystyle{abbrv-doi-hyperref}
%\bibliographystyle{abbrv-doi-hyperref-narrow}

%\bibliography{neurips_2023}

%%%%%%%%%%%%%%%%%%%%%%%%%%%%%%%%%%%%%%%%%%%%%%%%%%%%%%%%%%%%

\end{document}